\documentclass[12pt]{article}
\usepackage[utf8]{inputenc}
\usepackage{a4}
\usepackage{graphicx}
\usepackage{epstopdf}
\usepackage{xcolor}
\usepackage{mathtools}
\usepackage{cite}
\usepackage{lscape}
\usepackage{url}
\usepackage{authblk}

\begin{document}

\title{Surrogate Modeling of the CLIC Final-Focus System using Artificial Neural Networks}
\author[1]{J. \"{O}gren}
\author[1,2]{C. Gohil}
\author[1]{D. Schulte}
\affil[1]{European Organization for Nuclear Research (CERN), Geneva, Switzerland}
\affil[2]{John Adams Institute, University of Oxford, Oxford, United Kingdom}
\date{\today}
\maketitle
%
%
%
%%%%%%%%%%%%%%%%%%%%%%%%%%%%%%%%%%%%%%%%%%%%%%%%%%%%%%%%%%%%%%%%%%%%%%%
\begin{abstract}\noindent
Artificial neural networks can be used for creating surrogate models that can replace computationally expensive simulations. In this paper, a surrogate model was created for a subset of the Compact Linear Collider (CLIC) final-focus system. By training on simulation data, we created a model that maps sextupole offsets to luminosity and beam sizes, thus replacing computationally intensive tracking and beam-beam simulations. This model was then used for optimizing the parameters of a random walk procedure for sextupole alignment.
\end{abstract}
%
%%%%%%%%%%%%%%%%%%%%%%%%%%%%%%%%%%%%%%%%%%%%%%%%%%%%%%%%%%%%%%%%%%%%%%%
%
%
%						INTRODUCTION
%
%
%
%%%%%%%%%%%%%%%%%%%%%%%%%%%%%%%%%%%%%%%%%%%%%%%%%%%%%%%%%%%%%%%%%%%%%%%
\section{Introduction}
The field of machine learning has grown immensely in the past decade and is now finding its way into a wide range of fields. Particle accelerators are controllable physical systems with rich complexities and offer interesting applications that can profit from different machine learning methods. Examples of applications in particle accelerators are anomaly detection in superconducting magnets~\cite{wielgosz}, automatic control of collimators~\cite{azzopardi}, electron bunch profile prediction~\cite{emma,scheinker_prab2015}, electron beam size prediction in circular accelerators\cite{leemann} and ion source prediction~\cite{kong}. Different advanced techniques have also been used for optimzation and automatic control in free-electron lasers~\cite{bruchon,scheinker_prab2019}, longitudinal phase space control~\cite{scheinker_prl}, dynamic aperture optimization~\cite{Li} and optics correction~\cite{Li_prab19,fol}. A more comprehensive list of machine learning applications in particle accelerators can be found in~\cite{edelen_arxiv}.
\par
The field of particle accelerators relies to a large extent, especially in design and optimization, on high-fidelity simulations and there is an abundance of simulation tools. In many cases simulations are computationally expensive, which makes optimization time consuming and online modeling problematic. A surrogate model is an approximative model that can mimic the system and is particularly useful for applications that require extensive simulations. The surrogate model, which is fast to evaluate, can be used for optimization or as an online model. In this paper we make use of artificial neural networks and supervised deep learning to create a surrogate model for a part of the Compact Linear Collider (CLIC) with computationally expensive simulations: the final-focus system.
%
%
%%%%%%%%%%%%%%%%%%%%%%%%%%%%%%%%%%%%%%%%%%%%%%%%%%%%%%%%%%%%%%%%%%%%%%%
%
%
%						CLIC FFS
%
%
%
%%%%%%%%%%%%%%%%%%%%%%%%%%%%%%%%%%%%%%%%%%%%%%%%%%%%%%%%%%%%%%%%%%%%%%%
\section{The CLIC Final-Focus System}
CLIC is a proposed linear electron--positron collider~\cite{cliccdr,clicupd,clicpip} at CERN. To produce high luminosities, linear colliders rely on ultra-small beam sizes at the interaction point. This requires emittance preservation along the whole machine and puts tight tolerances on alignment and other imperfections. Being able to reach nominal performance under imperfect conditions is crucial for the reliability of such a machine. Nominal parameters for the 380~GeV energy stage are summarized in Table~\ref{tab:beam_params}.
\par
\begin{figure}%[h!]
\centering
\includegraphics[height=0.5\textwidth]{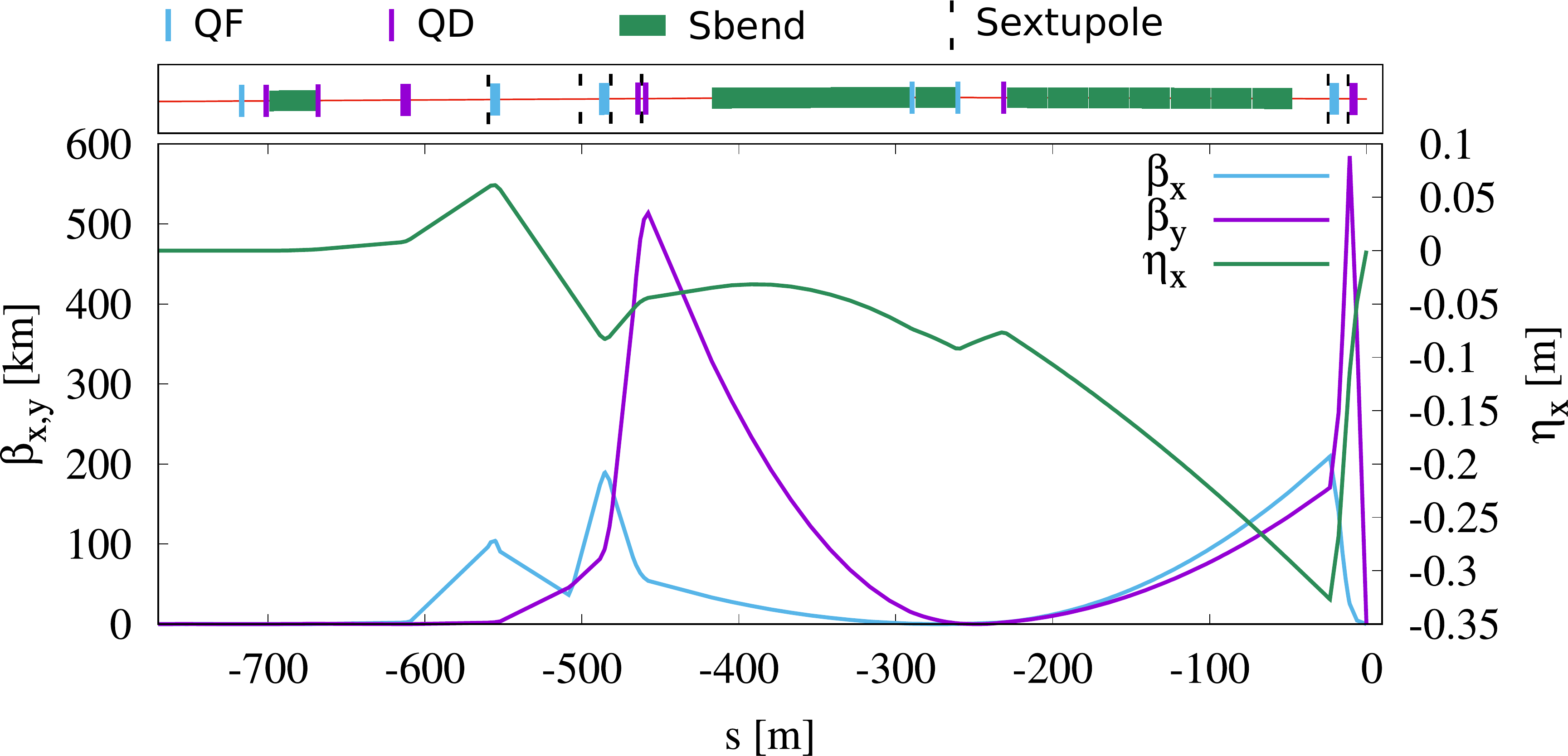}
\caption{\label{fig:ffs_optics}Optical functions for the 380~GeV CLIC final-focus system with $L^{*}$~=~6~m. Six sextupoles are placed in dispersive sections in a local chromaticity correction scheme.}
\end{figure}

The final-focus system constitutes the final 780~m of each beamline, see Figure~\ref{fig:ffs_optics}. Two strong quadrupole magnets at the end of the system de-magnify the beam in both transverse directions. Since particles of different energies are bent differently in a magnetic field, a beam with an energy spread will not be focused to a single point. This effect is called chromaticity and without compensation the beam size at the interaction point will be greatly inflated with diminished luminosity as a result. The chromaticity correction involves nonlinear magnets and a delicate compensation of unwanted aberrations. Complex systems such as this require fine-tuning and are challenging to design.
\par
The current version of the CLIC final-focus system has the final quadrupole doublet mounted outside the detector~\cite{clicdet,fab2018} with $L^{*}=6$~m---distance between the final quadrupole and point of collision. The system consists of 20 quadrupole magnets, 6 sextupole magnets and 2 octupole magnets. Bending magnets generate dispersion and in combination with the sextupole magnets, an energy-dependent focusing is generated that compensates the chromaticity generated in the final doublet. The final-focus system follows a local chromaticity correction scheme~\cite{local_ffs} where two sextupoles are placed at the final doublet and then additional sextupoles are placed upstream with optics designed in such a way that unwanted geometric aberrations cancel. Figure~\ref{fig:ffs_optics} shows the beta functions ($\beta_{x,y}$) and the dispersion profile ($\eta_x$) of the final-focus system.
%
%
% TABLE
\begin{table}
\centering
\caption{\label{tab:beam_params}380~GeV CLIC beam parameters.}
\begin{tabular}{lcc}
\hline
\hline
{\bf Parameter} & {\bf unit} & {\bf value} \\
\hline
Norm. emittance (end of linac) $\gamma\epsilon _{x}/ \gamma\epsilon _{y}$ & [nm] & 900 / 20 \\
Norm. emittance (IP) $\gamma\epsilon _{x}/ \gamma\epsilon _{y}$ & [nm] & 950 / 30 \\
Beta function (IP) $\beta_{x}^{*}/ \beta_{y}^{*}$ & [mm] & 8.2 / 0.1 \\
Target IP beam size $\sigma_{x}^{*}/ \sigma_{y}^{*}$ & [nm] & 149 / 2.9 \\
Bunch length $\sigma_{z}$ & [$\mu$m] & 70 \\
rms energy spread $\delta_{p}$ & [\%] & 0.35 \\
Bunch population $N_{e}$ & [$10^{9}$] & 5.2 \\
Number of bunches $n_{\mathrm{b}}$ & & 352 \\
Repetition rate $f_{\mathrm{rep}}$ & [Hz] & 50 \\
Luminosity $L$ & [$10^{34} \mathrm{cm}^{-2}\mathrm{s}^{-1}$] & 1.5 \\
Peak luminosity $L_p$ & [$10^{34} \mathrm{cm}^{-2}\mathrm{s}^{-1}$] & 0.9 \\
\hline
\hline
\end{tabular}
\end{table}
%
% FIGURE
%
%
%%%%%%%%%%%%%%%
\subsection{Tuning}
The final-focus system is sensitive to imperfections and the challenge is to achieve small beam sizes, and high luminosity, for a system with imperfections. Successful operation of a linear collider requires effective and fast tuning. Tuning studies have been done extensively for all parts of CLIC and for the final-focus system in particular. A recent report~\cite{statictuning} considered single-beam tuning of the 380~GeV CLIC final-focus system including static imperfections such as magnet offsets, roll errors and strength errors. Single beam means that only one side, i.e. one of the two beamlines, of the final-focus system is tracked and the beam at the interaction point is mirrored for the beam--beam simulation.
\par
In that study, the tuning procedure consisted of beam-based alignment that corrected the linear system, followed by sextupole alignment and tuning with sextupole and octupole knobs. Due to the nonlinearities of the system together with synchrotron radiation, small imperfections can have dramatic impact on beam size at the interaction point and consequently a substantial decrease in luminosity. The purpose of the tuning simulations is to find robust algorithms for achieving luminosity under imperfect conditions and thus ensuring the reliability of the collider. In all the simulations we use \texttt{PLACET}~\cite{placet} for tracking the beam through the final-focus system lattice and \texttt{GUINEA-PIG}~\cite{guineapig} for a full beam--beam simulation.
%
%%%%%%%%%%%%%%%
\subsection{Sextupole misalignment}
\begin{figure}%[h!]
\centering
\includegraphics[width=0.5\textwidth]{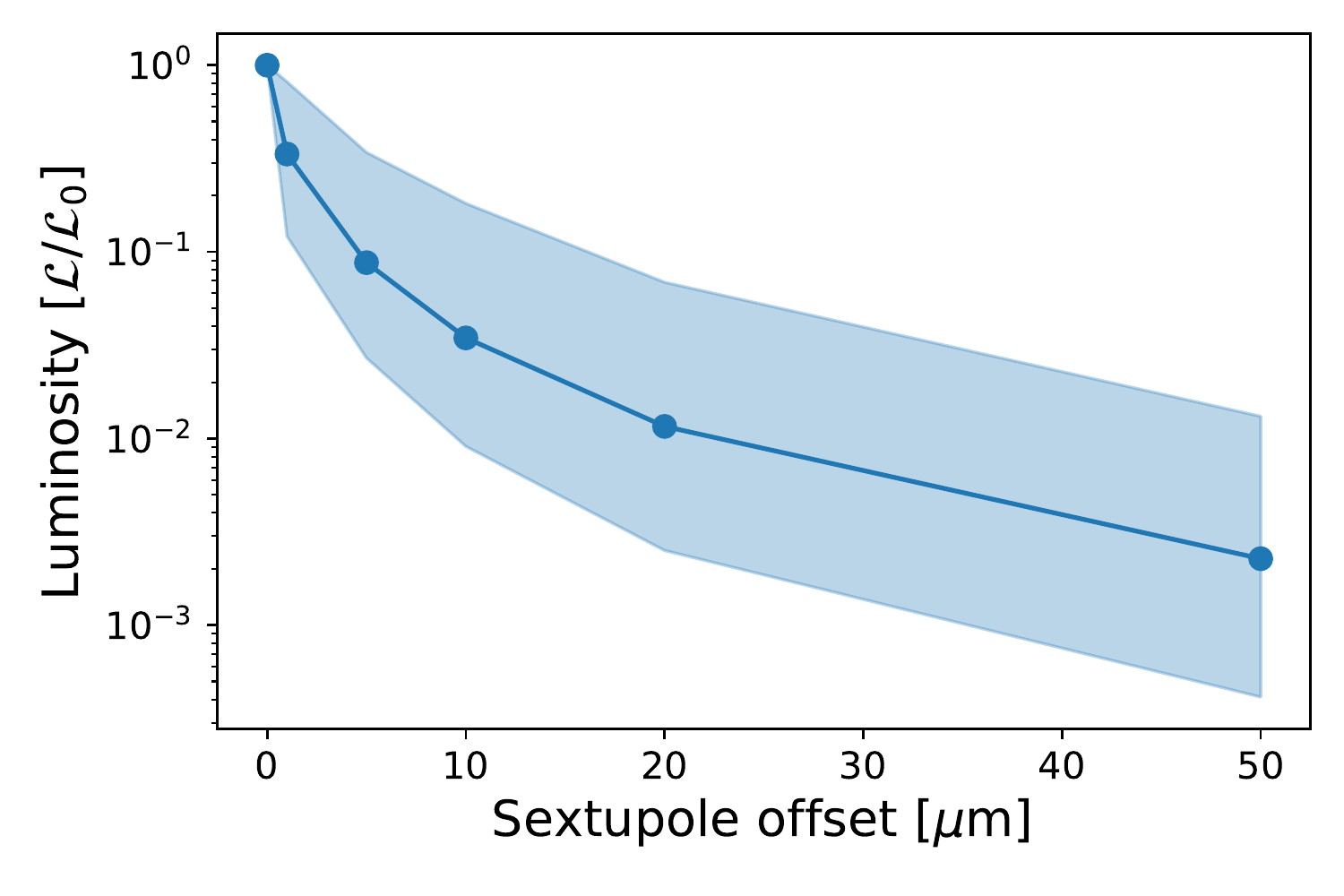}
\caption{\label{fig:missext}The average luminosity for the perfect machine with sextupole offsets of varying rms values. As the sextupole misalignments grow, luminosity quickly deteriorates. The shaded area show the maximum and minimum values.}
\end{figure}

Let us consider a single imperfection: transverse misalignment of sextupoles. Experience has shown that this imperfection has a large impact on luminosity. To quantify the impact we run simulations on a perfect machine and add transverse sextupole offsets assuming a normal distribution. For different rms values, we track 20 cases and run a beam--beam simulation to compute the luminosity. Similar to Ref.~\cite{statictuning} we consider the single-beam case. Figure~\ref{fig:missext} shows the average luminosity decreasing as the rms value of the transverse sextupole offset increases. Since transverse sextupole misalignments have a large impact, it is essential to have a quick and robust method for tuning the sextupoles. In the next section we build a surrogate model for the impact of sextupole misalignments on luminosity and beam size at the interaction point.
%
% FIGURE
%
%
%%%%%%%%%%%%%%%%%%%%%%%%%%%%%%%%%%%%%%%%%%%%%%%%%%%%%%%%%%%%%%%%%%%%%%%
%
%
%						SURROGATE MODEL
%
%
%
%%%%%%%%%%%%%%%%%%%%%%%%%%%%%%%%%%%%%%%%%%%%%%%%%%%%%%%%%%%%%%%%%%%%%%%
\section{Surrogate Model}
We only consider transverse sextupole offsets and aim to develop a model that outputs: luminosity, peak luminosity, horizontal beam size and vertical beam size. The peak luminosity is the luminosity of collisions within 1\% of the nominal energy. CLIC specifications require a minimum of 60\% of collisions occur at the nominal energy, see Table~\ref{tab:beam_params}.
\par
An analytical approach is not feasible due to the level of complexity of the system. The nonlinear fields of the sextupole magnets together with effects of synchrotron radiation make it difficult to compute the resulting beam distribution at the interaction point. Furthermore, the disruption parameter in the beam--beam interaction is high, which means that the beams have considerable effect on each other. In an electron--positron collider the transverse electromagnetic fields of one beam focuses the opposing beam (`pinch effect'). In such circumstances analytical estimates of luminosity are not reliable.
\par
Artificial neural networks are large connections of small units (artificial neurons) and by training on data the models can learn to perform different tasks such as classification, clustering or regression. In our case we make use of regression and train a model to map input to known outputs. In the context of machine learning, this is known as supervised learning. We use the machine learning system TensorFlow~\cite{tensorflow} interfaced via the Python library Keras~\cite{keras} for training the models. The goal of a model is to make accurate predictions based on unseen data. Such models are said to generalize well. Large data sets are often needed to train artificial neural networks successfully.
%
%%%%%%%%%%%%%%%
\subsection{Data generation}
\begin{figure}%[h!]
\centering
\includegraphics[width=1.0\textwidth]{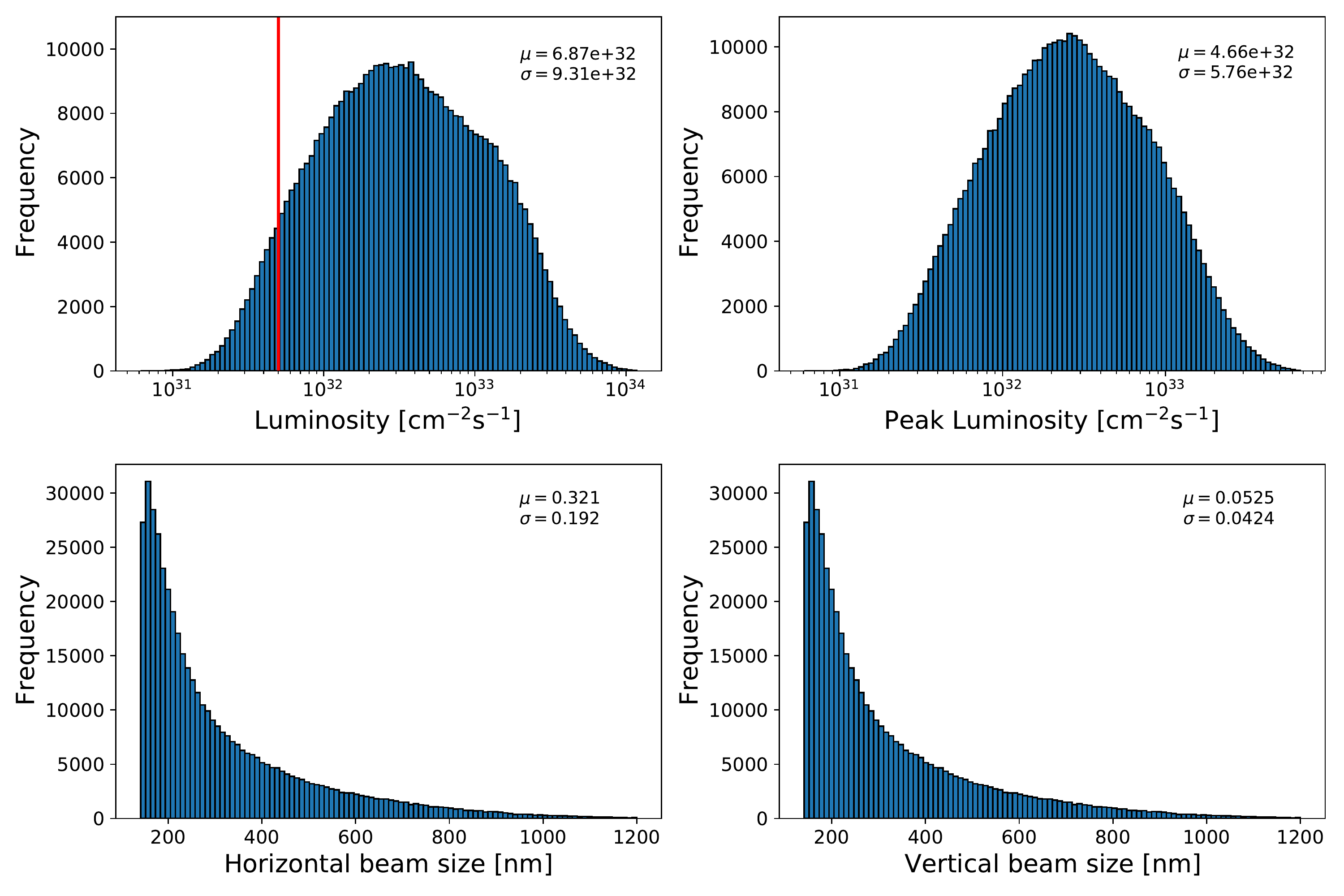}
\caption{\label{fig:sextdata}Histograms of the data set showing the four outputs: luminosity (top left), peak luminosity (top right), horizontal beam size (bottom left) and vertical beam size (bottom right). The vertical line in the luminosity plot shows a cut that was used before training the models.}
\end{figure}

Assuming normal distributions with rms values of 5, 10 and 20~$\mu$m, we added random offsets to the sextupoles. For each case we tracked $10^5$ macroparticles through the final-focus system, ran a beam-beam simulation and saved sextupoles positions, beam sizes at the interaction point and the output from the beam-beam simulation: luminosity and peak luminosity. The data was generated in parallel using the CERN batch system and in a few days we had accumulated a data set of 450,000 cases. Each case took approximately 5 minutes to run on an ordinary desktop computer (8 cores, CPU computations only). Figure~\ref{fig:sextdata} shows histograms of the accumulated data. We use 100 bins in the histogram because we found that this was sufficient to resolve the shape of the distribution.
\par
\begin{figure}%[h!]
\centering
\includegraphics[width=1.0\textwidth]{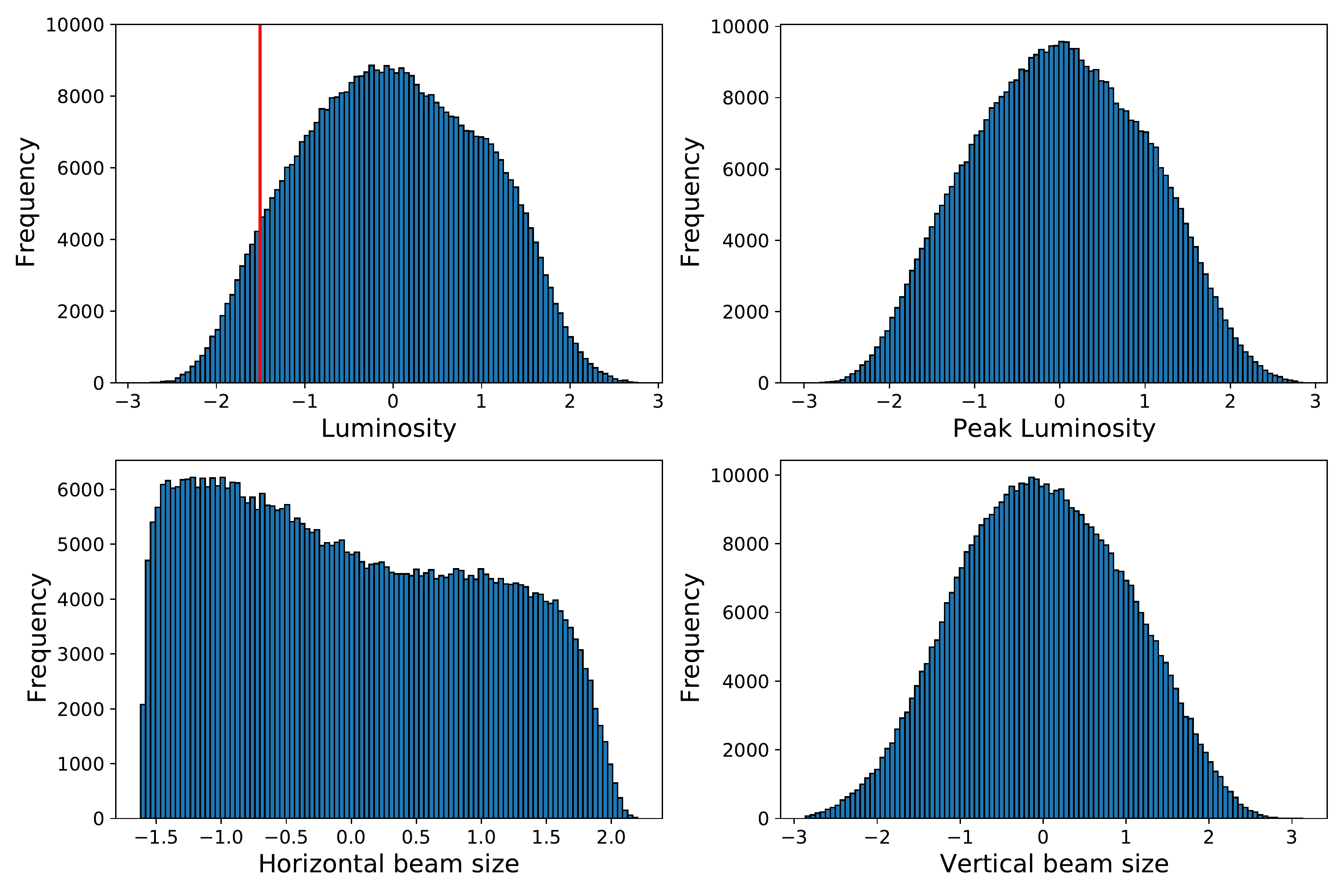}
\caption{\label{fig:sextdata_pt}Histograms of the data set after power transformation and standardization: luminosity (top left), peak luminosity (top right), horizontal beam size (bottom left) and vertical beam size (bottom right). The data is also standardized to have zero mean and unit variance. The vertical line in the luminosity plot shows a cut that was used before training the models.}
\end{figure}

Since the numeric values of the inputs and outputs have very different ranges we must transform data before training our models. We standardize the data set, i.e. apply transformations $X_{standard} = (X-\mu)/\sigma$, so that each column in our data set has zero mean and unit variance. The mean $\mu$ and standard deviation $\sigma$ are computed using the training data set. To the columns containing output data (luminosity, peak luminosity, horizontal and vertical beam sizes) we also apply a power transformation to make distributions closer to Gaussian, see Fig.~\ref{fig:sextdata_pt}. Standardization and power transformation are common practices to make the training of artificial neural networks numerically stable. Both these transformations are available in the Python library Scikit-learn~\cite{scikit}. When testing the model, the inverse transformations are applied to the output predictions.
%
% FIGURE
%
% FIGURE
%
%
%%%%%%%%%%%%%%%
\subsection{Model setup}
There are no clear guidelines of how to set up an artificial neural network and how to select a model architecture since it depends on the problem. In general, if the model has a complexity that is too low it will not be able to mimic nonlinear behavior and if the model has too much complexity it can suffer from overfitting, where the model does not generalize well.
\par
\begin{figure}%[h!]
\centering
\includegraphics[width=0.5\textwidth]{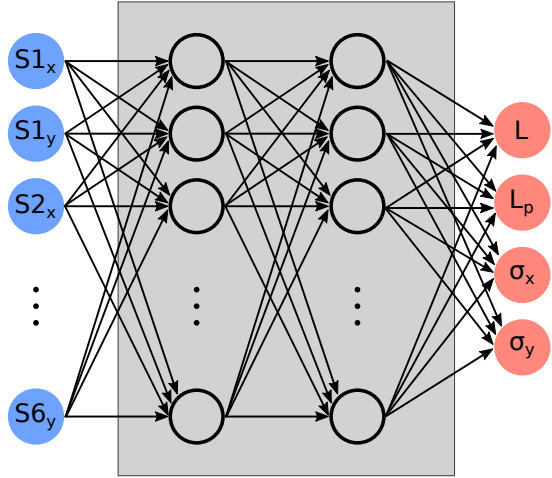}
\caption{\label{fig:model}A schematic of an artificial neural network model that maps the horizontal and vertical positions of the six sextupole magnets in the CLIC final-focus system to the resulting luminosity, peak luminosity and horizontal and vertical beam sizes. This model consists of two fully connected hidden layers in a feedforward configuration.}
\end{figure}

First we consider a simple case were our model consists of two hidden layers of twelve neurons each. We use a model with fully connected layers in a feedforward configuration (every neuron in one layer connects to every neuron in the subsequent layer). Figure~\ref{fig:model} shows a schematic of the model. The 12-dimensional input vector can be written as
\begin{equation}
	X = (x_1, \dots, x_{12})^T = (S1_x, S1_y, \dots, S6_x, S6_y)^T
\end{equation}
where $S1_x$ denotes the horizontal position of the first sextupole and so on. The 4-dimensional output can be written
\begin{equation}
	Y = (y_1, \dots, y_4)^T = (L, L_p, \sigma_x, \sigma_y)^T
\end{equation}
where $L$ denotes the luminosity, $L_p$ the peak luminosity and $\sigma_x$ ($\sigma_y$) the horizontal (vertical) beam size at IP.
The input to a neuron is the sum of weighted inputs and a bias, and during the training of the model, weights and biases of all neurons are adjusted to minimize a loss function---in our case the mean squared error.
%
% FIGURE
%
%%%%%%%%%%%%%%%
\subsection{Model Training}
Following common practice, we split the complete data set and reserve 20\% for testing. The remaining 80\% is split into a training set (90\%) and a validation set (10\%). During training the model trains on the training set and evaluates on the validation set. The purpose of validation is to monitor overfitting. After training is done, i.e. when there is either no improvements or overfitting starts, the model performance is evaluated on the testing data---data never previously seen by the model.
\par
To assess the model performance we compare the predictions to the known outputs in the test data set. As a figure of merit we use the mean absolute percentage error (MAPE). From Figure~\ref{fig:sextdata} it is clear there are a few outliers with low luminosity. This results in large errors which can inflate the MAPE. Therefore, we make a cut in our testing data and only consider cases with luminosities above $5\times10^{31}$~cm$^{-2}$s$^{-1}$.
\par
\begin{figure}%[h!]
\centering
\includegraphics[width=1.0\textwidth]{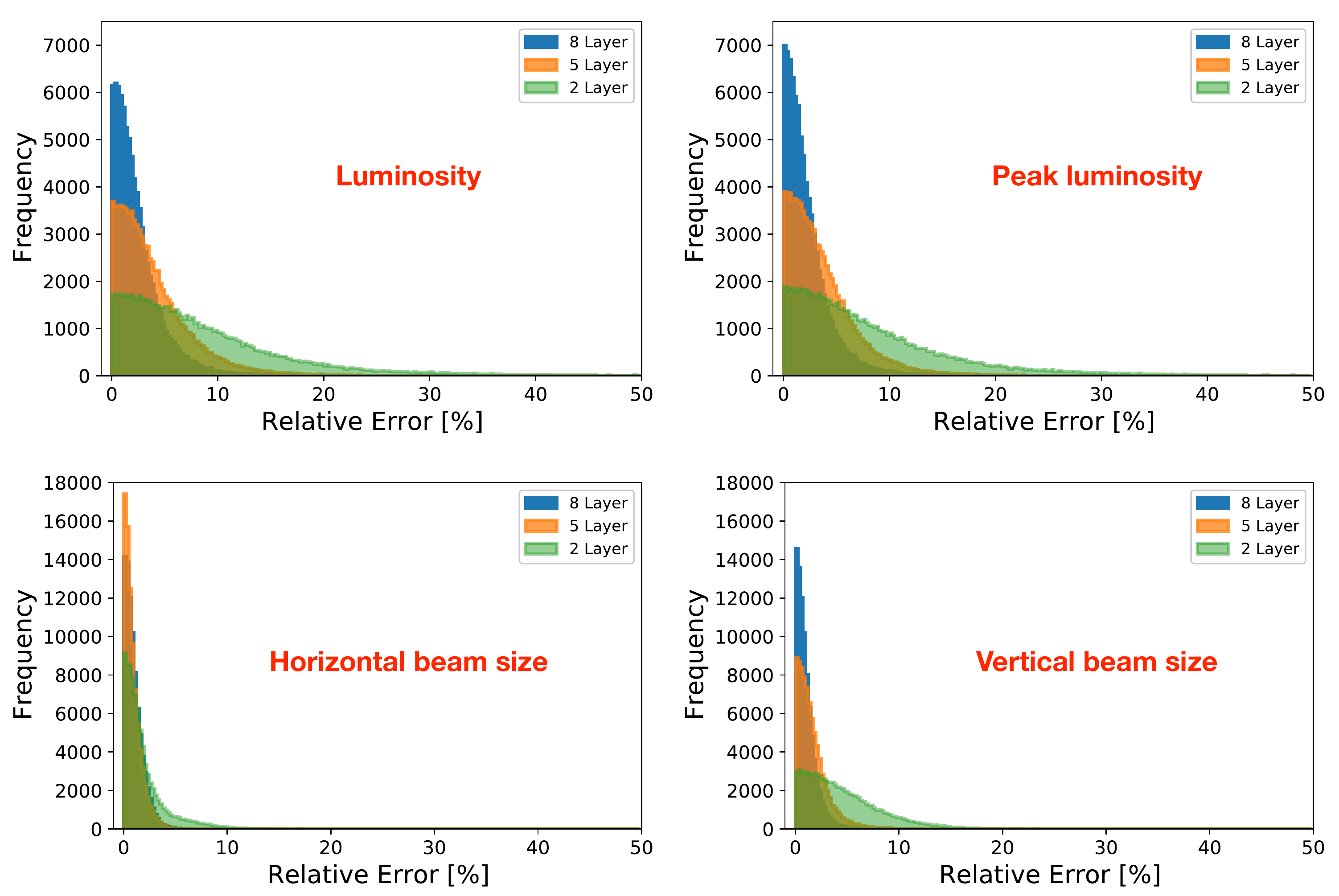}
\caption{\label{fig:evaluation}Histograms of the absolute percentage error for the four different outputs: luminosity (top left), peak luminosity (top right), horizontal beam size (bottom left) and vertical beam size (bottom right). Three different models with two, five and eight layers are compared.}
\end{figure}

Figure~\ref{fig:evaluation} shows histogram of the performance of the three different models. We tested a different number of hidden layers (2-9) and the plots show three cases: two, five and eight layers. It is clear that a model with two layers does not have sufficient complexity and gives rather large relative errors. There is a substantial improvement when we increase the number of layers from two to five and eight. Table~\ref{tab:model_eval} shows the mean of the absolute percentage error. Eight layers gave the lowest MAPE and there was no improvement from increasing the number of layers beyond eight. We also tried other network architectures, for example: we increased the number of nodes per layer; tried reducing the number of nodes in subsequent layers and tried architectures with a large number of nodes in the first layer followed by smaller layers. In the end it did not seem possible to lower the MAPE further.
%
% FIGURE
%
% TABLE
\begin{table}
\centering
\caption{\label{tab:model_eval}Model evaluation on testing data. Mean absolute percentage error for the four model outputs for three different model architectures.}
\begin{tabular}{l|ccc}
\hline
\hline
& 2 layers & 5 layers & 8 layers \\
\hline
Luminosity [\%] 			& 8.87 & 4.01 & 2.47 \\
Peak luminosity [\%]		& 8.42 & 3.77 & 2.29 \\
Horizontal beam size [\%] 	& 2.06 & 0.913 & 1.02 \\
Vertical beam size [\%]		& 4.74 & 1.68 & 1.04 \\
\hline
\hline
\end{tabular}
\end{table}
%
%
%%%%%%%%%%%%%%%
\subsection{How much data is needed?}
\begin{figure}%[h!]
\centering
\includegraphics[width=0.6\textwidth]{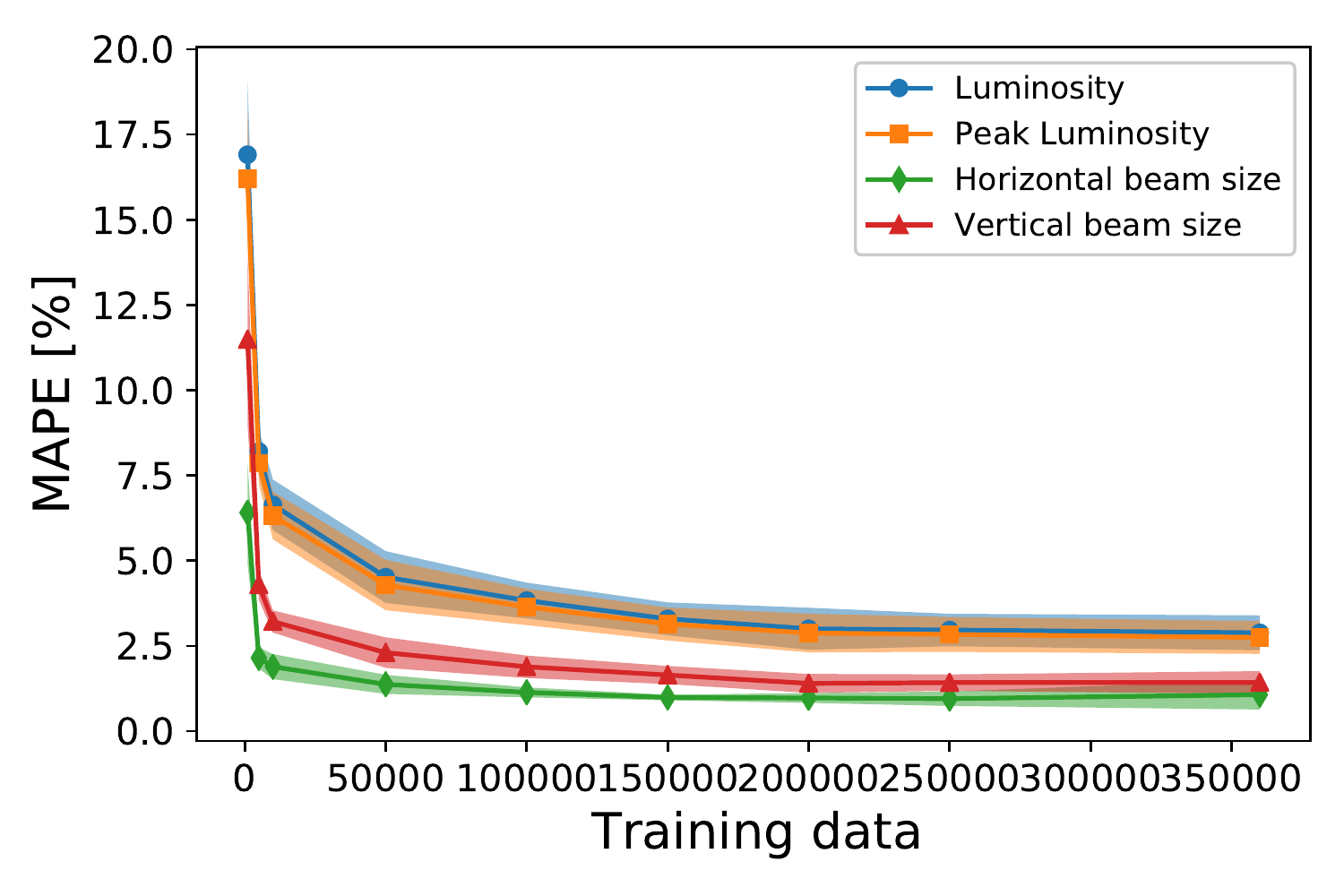}
\caption{\label{fig:mape_vs_size}Mean absolute percentage error (MAPE) for models trained on different number of training samples. Each case is trained ten times and the markers show the mean values and the shaded areas the standard deviation. It is clear that more data is better but with diminishing returns.}
\end{figure}

The full data set used in the previous section consisted of 450,000 samples containing misalignments drawn from a random distribution. 360,000 samples were randomly selected for the training dataset and the remaining 90,000 samples were used for testing. To investigate the influence of number of data points, we train the model with eight layers on training sets of different sizes. Subsets of [1000, 5000, 50000, 100000, ..., 300000] are drawn randomly from the full training set. Each case is trained ten times and tested on the same full testing set used in previous section. Figure~\ref{fig:mape_vs_size} shows the mean and standard deviation of the MAPE for the models trained on different number of data points. It is clear that more data is better but with diminishing returns. Already at 50,000 data points the model performance approaches the final result. For a model valid for luminosities above $5\times10^{31}$~cm$^{-2}$s$^{-1}$, 50,000 data points seems sufficient. On the other hand, in order to increase the range of the model more data and more data over a wider range is likely required.
%
% FIGURE
%
%
%
%%%%%%%%%%%%%%%
\subsection{Final Model}
The final model consisted of eight fully connected layers in a feedforward configuration, c.f. Figure~\ref{fig:model}. Each layer had 12 neurons that used the rectified linear unit as an activation function. The model was trained on the full training set of 360,000 data points using the Adam optimizer~\cite{tensorflow}---an adaptive learning method implemented in TensorFlow---and mean squared error as loss function. We used a batch size (number of training samples used in every iteration) of 50 and trained for 400 epochs (the number of times the complete training data set is passed through). The final model took a total of 8-10 hours to train on an ordinary desktop computer (8 cores, CPU computation). The results from testing are presented in Table~\ref{tab:model_eval}.
\par
Surely there is potential room for improvement. For instance, the hyperparameter optimization was not done exhaustively and other model architectures might be better. However, finding the optimum model is not within the scope of this paper. Instead, we want to show that the model was good enough to be useful. To assess the final model's usability we compared it to full simulations.
%
%%%%%%%%%%%%%%%%%%%%%%%%%%%%%%%%%%%%%%%%%%%%%%%%%%%%%%%%%%%%%%%%%%%%%%%
%
%
%						MODEL VERIFICATION
%
%
%
%%%%%%%%%%%%%%%%%%%%%%%%%%%%%%%%%%%%%%%%%%%%%%%%%%%%%%%%%%%%%%%%%%%%%%%
\section{Model Verification}
To verify the model performance we compare it to the full simulation. A nice feature with the tracking code \texttt{PLACET} is it can easily interface Python. This made it possible to seamlessly load and interface the trained machine learning model in our simulations. At each iteration we ran the full beam--beam simulation and at the same time checked the prediction of the machine learning model for comparison.
\par
\begin{figure}%[h!]
\centering
\includegraphics[width=\textwidth]{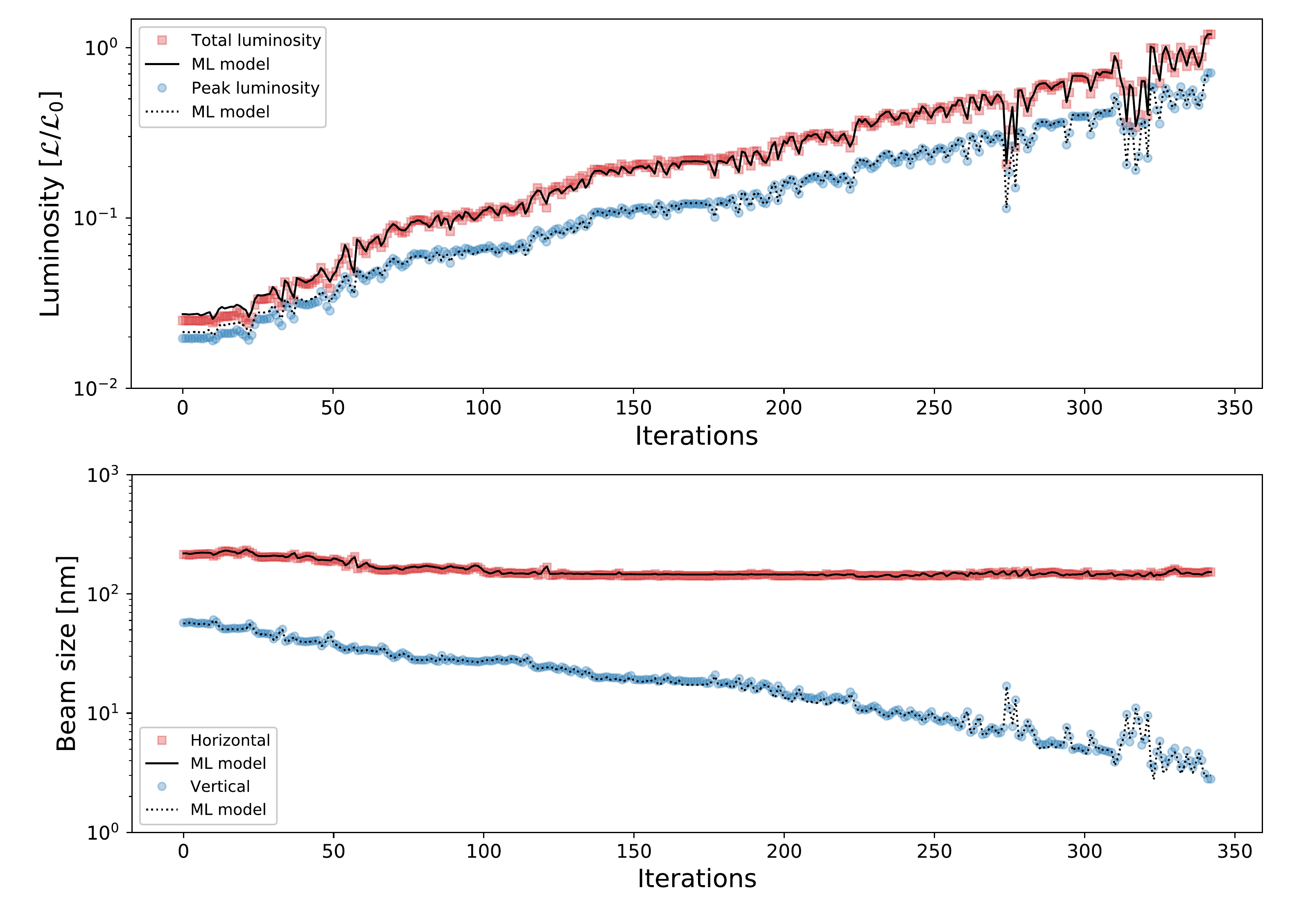}
\caption{\label{fig:compare}Example of a random walk optimization on the full simulation.}
\end{figure}

Figure~\ref{fig:compare} shows a random walk tuning of the transverse sextupole position. Random sextupole offsets were added to a perfect machine and using a random walk the luminosity was optimized within 350 iterations. At each iteration the beam distribution was tracked and the luminosity was computed with a full beam--beam simulation. At each step we also checked the prediction of the machine learning model (plotted in solid and dotted lines). We plot luminosity normalized to nominal luminosity, $L_0$, which is the luminosity of the unperturbed lattice. Note that sometimes the luminosity in the random walk tuning is higher than the nominal value since sextupole offsets can introduce waist shifts that enhances luminosity via the pinch effect and the nominal luminosity does not include this effect. Over the range of almost two orders of magnitude there was excellent agreement between the machine learning model and the full simulation. The MAPE in this case was 2.6\% for luminosity, 3.0\% for peak luminosity, 1.8\% for horizontal beam size and 2.6\% for vertical beam size. The MAPE was higher in this simulation compared to the testing dataset, which showed a MAPE of approximately 2\% for luminosity and 1\% for beam size. This is due to the small sample size in this simulation which leads to a higher variance in the MAPE calculated. The random walk method always used the luminosity from the full simulation and the machine learning model was only evaluated parasitically. 
\par
The converse was also tested, where the output of the machine learning model guided the random walk optimization and the full simulation was done parasitically for comparison. Both cases showed similar results.
%
% FIGURE
%
%
%%%%%%%%%%%%%%%%%%%%%%%%%%%%%%%%%%%%%%%%%%%%%%%%%%%%%%%%%%%%%%%%%%%%%%%
%
%
%						RANDOM WALK OPT
%
%
%
 %%%%%%%%%%%%%%%%%%%%%%%%%%%%%%%%%%%%%%%%%%%%%%%%%%%%%%%%%%%%%%%%%%%%%%%
\section{Random Walk Optimization}
Running a single case on the full simulation, i.e. tracking $10^5$ macroparticles and running the full beam--beam simulation, on a normal desktop computer takes a few minutes. Making a single prediction with the machine learning model takes a few milliseconds. Such a speed-up in performance allows for systematic studies of optimization otherwise not feasible. As an example, we use our machine learning model to optimize a simple random walk procedure.
\par
For the sextupoles, there are twelve degrees of freedom: six sextupoles and two transverse directions. We devise a random walk procedure in the following way. Out of the twelve degrees of freedom a subset of the degrees of freedom is selected randomly. In this subset a direction is selected at random, i.e. we specify a random linear combination of the degrees of freedom in the subset. Four points along the selected direction, $g[-1,-0.5,0.5,1]$ with gain $g$, are checked. If the highest luminosity of the four points exceeds the luminosity from the previous iteration, that points is selected. Otherwise the procedure moves to the next iteration and selects a new subset and direction. The whole procedure is iterated until a certain luminosity is reached. There are two parameters to be optimized: gain and size of the subset.
\par
\begin{figure}%[h!]
\centering
\includegraphics[width=0.5\textwidth]{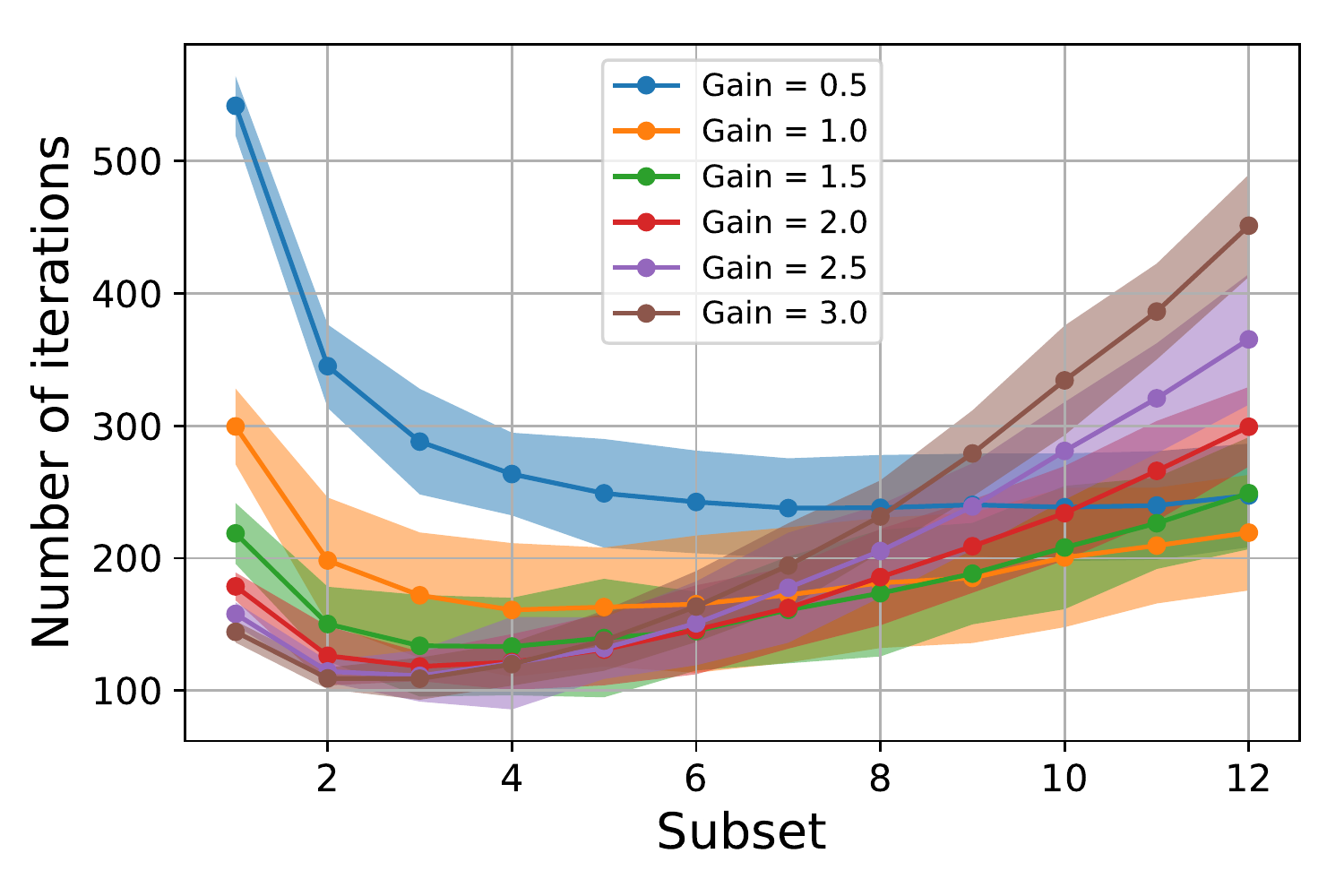}
\caption{\label{fig:RW_opt}Grid search of gain and subset used in a random walk procedure. The points show the average values and shaded area the standard deviation. For each setting a 100 random cases were tested 100 times.}
\end{figure}

We performed a grid search of gains $[0.5, 1.0, 1.5, 2.0, 2.5, 3.0]$ and subsets 1-12 on the eight layer artificial neural network model. For each setting we tested 100 different initial conditions with random sextupole offsets. For each initial condition we ran the random walk procedure 100 times and recorded the required number of iterations to reach the target luminosity. Figure~\ref{fig:RW_opt} shows the average number of iterations needed for the different parameter settings. For a smaller subset a higher gain decreases the number of iterations whereas for a larger subset a smaller gain is preferable. The optimum seems to be a small subset of 2 or 3 and a higher gain.
%
% FIGURE
\par
\begin{figure}%[h!]
\centering
\includegraphics[width=0.8\textwidth]{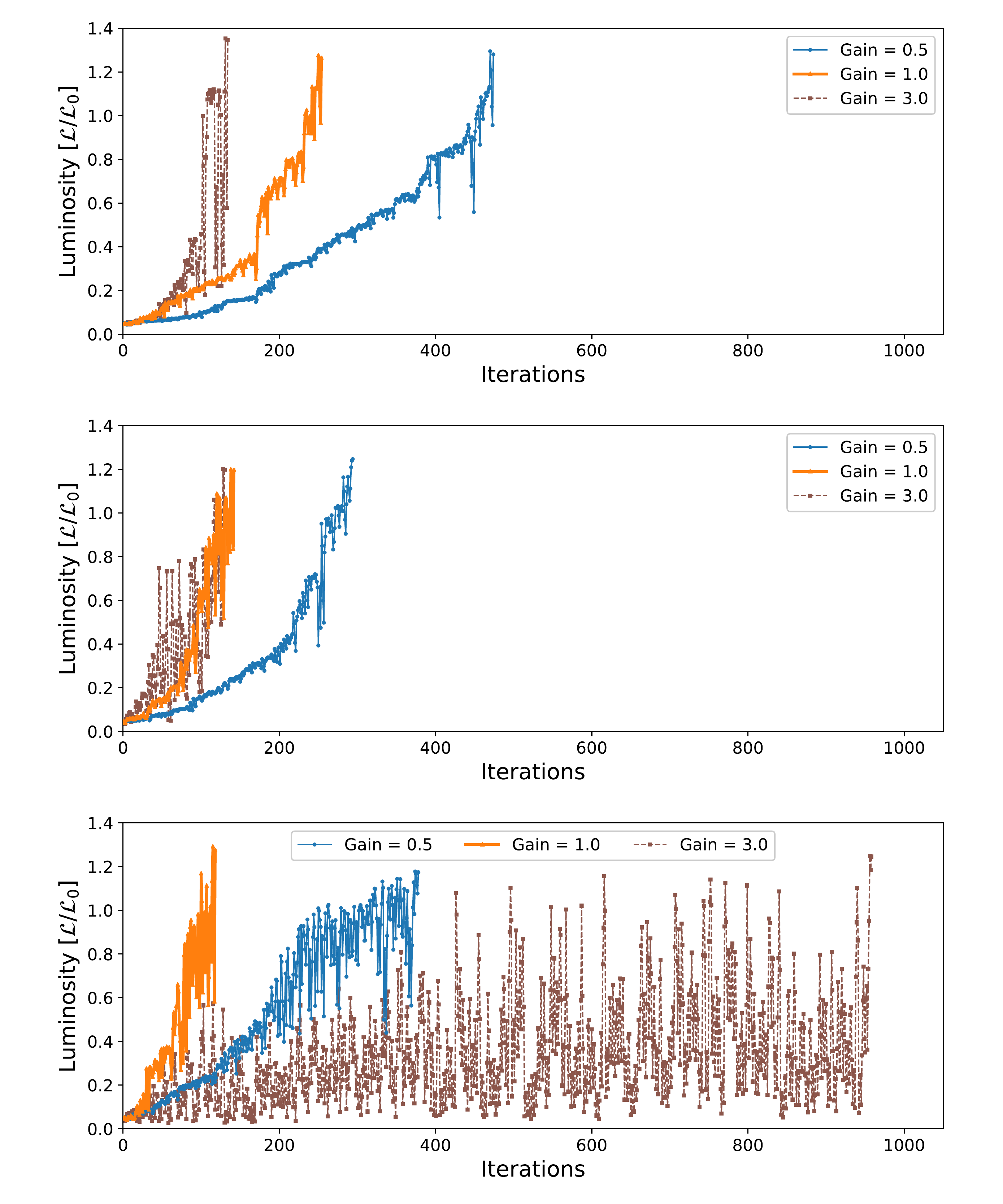}
\caption{\label{fig:RW_check}Random walk procedure tested on the full simulation. Three different gains were tested using subset = 2 (top), subset = 6 (middled) and subset = 12 (bottom). A higher gain is better when the subset is small and a smaller gain when subset is large. }
\end{figure}

To verify the optimum parameters from the grid search, we tested the random walk procedure using a full simulation. Figure~\ref{fig:RW_check} shows three different cases: subsets 2, 6 and 12. For each case, three different gains were tested. The results are in good agreement with the grid search using the machine learning model. Using a higher gain is better when the subset is small whereas a high gain and a large subset lead to large fluctuations and requires more iterations to reach target.
%
% FIGURE
\par
This random walk procedure was rather simple and could surely be optimized further. In a real machine there are other constraints to consider as well and in most cases a slow smoothly varying optimization might be preferred over a quicker procedure with large fluctuations. The point of this exercise was to show the usability of the surrogate model and that results from optimizations on the surrogate model translated to the full simulation.
%
%%%%%%%%%%%%%%%%%%%%%%%%%%%%%%%%%%%%%%%%%%%%%%%%%%%%%%%%%%%%%%%%%%%%%%%
%
%
%						CONCLUSIONS
%
%
%
%%%%%%%%%%%%%%%%%%%%%%%%%%%%%%%%%%%%%%%%%%%%%%%%%%%%%%%%%%%%%%%%%%%%%%%
\section{Conclusions}
In this paper we presented a surrogate model that maps sextupole offsets to resulting luminosities and beam sizes in the CLIC final-focus system. We trained an artificial neural network on a large set of simulation data. The model was verified by comparing its predictions to the full simulation with a mean absolute percentage error of a few percent. We used the model to optimize two parameters of a random walk procedure and compared the results to the full simulation, with good agreement.
\par
We considered only a small subset of the system (transverse sextupole offsets) and the single-beam case. A more complete surrogate model could be realized by adding more imperfections to the system. Naturally, as the degrees of freedom increase, training the model becomes more challenging and may require a larger data set but the principle remains the same. Nonetheless, as we showed with our example of optimizing the random walk procedure: even a model with limited scope can already be very useful. In our case it helped improving the sextupole alignment procedure and reduced the overall tuning time.
%
%
%%%%%%%%%%%%%%%%%%%%%%%%%%%%%%%%%%%%%%%%%%%%%%%%%%%%%%%%%%%%%%%%%%%%%%%

\end{document}